# Topological semimetal state and field-induced Fermi surface reconstruction in antiferromagnetic monopnictide NdSb

Yongjian Wang,[1,2] J. H. Yu,[3] Y. Q. Wang,[1] C. Y. Xi,[1] L. S. Ling,[1] S. L. Zhang,[1] J. R. Wang,[1] Y. M. Xiong,[1] Tao Han,[1] Hui Han,[1] Jun Yang,[1] Jixiang Gong,[1] Lei Luo,[1] W. Tong,[1] Lei Zhang,[1] Zhe Qu,[1] Y. Y. Han,[1] W. K. Zhu,[1,4,*] Li Pi,[1,2,4,†] X. G. Wan,[3,4] Changjin Zhang,[1,4,‡] and Yuheng Zhang[1,2,4]

[1]*Anhui Province Key Laboratory of Condensed Matter Physics at Extreme Conditions, High Magnetic Field Laboratory, Chinese Academy of Sciences, Hefei 230031, China*

[2]*Hefei National Laboratory for Physical Sciences at Microscale, University of Science and Technology of China, Hefei 230026, China*

[3]*National Laboratory of Solid State Microstructures, College of Physics, Nanjing University, Nanjing 210093, China*

[4]*Collaborative Innovation Center of Advanced Microstructures, Nanjing University, Nanjing 210093, China*

## Abstract

We report the experimental realization of Dirac semimetal state in NdSb, a material with antiferromagnetic ground state. The occurrence of topological semimetal state has been well supported by our band structure calculations and the experimental observation of chiral anomaly induced negative magnetoresistance. A field-induced Fermi surface reconstruction is observed, in response to the change of spin polarization. The observation of topological semimetal state in a magnetic material provides an opportunity to investigate the magneto-topological phenomena.

# I. INTRODUCTION

The three-dimensional (3D) topological semimetal (TSM) state not only provides a fascinating bridge between high energy and condensed matter physics, but also serves as an ideal platform to investigate spontaneous symmetry breaking, phase transition and many other fundamental phenomena in nature [1-7]. In 3D topological Dirac semimetals, the Dirac points are protected by both time-reversal symmetry and crystalline inversion symmetry [3-5]. When inversion or time-reversal symmetry is broken, the Dirac point may split into pairs of Weyl points and the Dirac semimetal becomes a Weyl semimetal [7].

Recently, 3D Dirac semimetal state has been realized in several compounds such as $Cd_3As_2$ and $A_3Bi$ (A is Na, K, or Rb) [4,8-10]. Meanwhile, the low-energy Weyl fermion quasiparticle has been theoretically predicted in TaAs and related materials [11,12]. The fundamental experimental signatures of Weyl semimetals, such as the surface Fermi arcs and the chiral-anomaly-induced negative magnetoresistance, have been successfully observed [13-17]. A noticeable fact is that these TSMs are all discovered in nonmagnetic compounds. There are few reports on the experimental discovery of TSM in a material with magnetic ground state. As a matter of fact, the initial proposals of TSMs are involved in magnetic materials, where the spin degeneracy of the bands is removed by breaking time reversal symmetry [1,2]. Unfortunately, the experimental observation of TSM states in magnetic materials such as $Y_2Ir_2O_7$ and $HgCr_2Se_4$ is very difficult, mainly due to the absence of high-quality samples. Nevertheless, a new kind of Dirac semimetals, which are protected by combined crystalline symmetry and magnetic structure, have recently been predicted to exist [18-21]. Thus the possible topological semimetal state is awaiting realization in magnetic materials.

In this paper we report the discovery of TSM state in NdSb, a material with antiferromagnetic (AFM) ground state. The existence of Dirac semimetal state is well supported by our band structure calculations and the observation of chiral anomaly induced negative magnetoresistance (MR). Furthermore, a field-induced Fermi surface reconstruction is observed, when the magnetic structure is changed to be ferromagnetic (FM) under high magnetic field.

# II. EXPERIMENTAL DETAILS

Single crystals of NdSb were grown with Sn flux. It has a NaCl-type structure with space group

*Fm*-3*m*, sharing the same structure with LaSb [22]. Magnetic measurements were taken on a Quantum Design MPMS-3 and the Cell-2 Magnet of High Magnetic Field Laboratory, Chinese Academy of Sciences (CHMFL). The low-field magnetoresistance was measured on a Quantum Design PPMS. The high-field MR was measured on the Cell-1 Magnet of CHMFL. The electronic structures were calculated by first-principles density functional theory (DFT) using the WIEN2k code [23]. The Perdew-Burke-Ernzerhof functional [24] was employed with a method of LDA+*U* (*U*=6.4 eV, calculated according to Ref. [25]). The *k*-mesh consists of 8000 points and lattice constant *a*=6.319 Å.

## III. RESULTS AND DISCUSSION

Figure 1(a) shows the magnetoresistance (MR) $\frac{\Delta\rho}{\rho_0} = \frac{\rho_H - \rho_0}{\rho_0}$ of NdSb in the field range of ±8 T at different temperatures. The MR reaches up to 30000% at 2 K and 8 T, with no indication of saturation. When the applied magnetic field is larger than 2 T, clear Shubnikov-de Haas (SdH) oscillation signal can be detected. Figure 1(b) shows the Hall resistivity $\rho_{xy}$ of NdSb in the range of 0-8 T. Figure 1(c) presents the obtained oscillations at low temperatures as a function of reciprocal field. The fast Fourier transformation (FFT) spectra reveal several oscillation frequencies, i.e., 25 T ($\delta$), 29 T ($\varepsilon$) and 235 T ($\alpha$), as shown in Fig. 1(d). The magnetoresistance in low field is consistent with previous reports [26].

Figure 2(a) presents the magnetic field dependence of magnetization under different temperatures. A sharp increase in magnetization occurs at around 10 T. Then the magnetization climbs to a plateau that persists up to 24.5 T. The plateau at 1.8 K corresponds to a magnetic moment of 2.81 $\mu_B$ per formula unit, which is close to the saturated moment of $Nd^{3+}$, i.e., 3 $\mu_B$. Therefore, the transition is from AFM state to FM state. Figure 2(b) shows the details of the transition at 1.8 K, which confirms the transition nature. In view of this AFM-FM transition, we study the transport properties below and above the critical field ($H_C$), respectively.

Figure 2(c) shows the high-field MR data taken at various temperatures. The MR reaches up to $10^6$% at 1.4 K and 38 T. The MR curves reveal a distinct change in oscillation frequency in the field range of 8-11 T. For example, at 0.39 K the change in oscillation occurs at a $H_C$ of 9.6 T [Fig. 2(d)]. The critical fields determined from the magnetoresistance data [inset of Fig. 2(d)] is consistent with those from the magnetization data [inset of Fig. 2(b)].

In order to check the electronic structures, we perform the DFT calculations for NdSb. The spin polarization is set as type-I face-centered-cubic (fcc) AFM (along [111] direction) and FM, respectively. Figures 3(a)-3(c) show the Fermi surface topologies for the AFM state, and the FM state with spin up and spin down, respectively. In the AFM state, the Fermi surfaces are identical for spin up and spin down cases, which are thus degenerate. Two predominant Fermi surfaces are found in the Brillouin zone center (Γ point), and the other two are found in the zone corner, centered at the X point and K point, respectively [see Fig. 3(i) for the high symmetry points in Brillouin zone]. The situation becomes different in the FM state, where the Fermi surfaces are slightly different for the spin up and spin down cases, as shown in Figs. 3(b) and 3(c). The main difference between the AFM state and FM state is that the Fermi surface at K point present in the AFM state is absent in the FM state.

The band structures are further plotted along the *k*-path through high symmetry points. As shown in Fig. 3(d), the Fermi surfaces are characteristic of Fermi pockets, namely, hole-type pockets in the zone center and electron-type pockets in the zone corner. A noticeable feature is that a band inversion occurs near the X point in the U1-X line, about 0.37 eV lower than the Fermi level. For the FM state, we obtain the band structures in Figs. 3(e) and 3(f) for spin up and spin down, respectively. Two substantial differences from the AFM band structures can be found. First, the conduction band and valence band at K point get far away from each other and the electron pocket vanishes. This suggests a Fermi surface reconstruction during the transition from AFM to FM. Second, the band inversion near X point disappears and instead a band crossing occurs at X point.

In order to understand the nature of these crossing points, their degeneracies are checked within the U-X-W plane. The band structures are re-calculated along the U1-X-U2 path, where U2 is the adjacent equivalent point of U1 for the crystalline symmetry. The nonsymmetrical band structures in Fig. 3(g) indicate that there exists a pair of crossing points in the U-X-W plane for the AFM state, while the symmetrical band structures in Fig. 3(h) show only one crossing point for the FM state. Considering the degenerate spin-up and spin-down conditions for the AFM state, it is found that the crossing point close to X is a Dirac point. This is consistent with the predication of the existence of Dirac fermions in an AFM semimetal [18-21]. For the FM state, the crossing point at X is two-fold degenerate due to the presence of net magnetic moment.

The negative longitudinal magnetoresistance (LMR) induced by chiral anomaly is an important character of topological semimetal state, which has been observed in a number of TSMs, including $Cd_3As_2$, $Na_3Bi$, $ZrTe_5$, TaAs, $WTe_2$, GdPtBi [16,27-31], etc. In Dirac semimetals, the Dirac point can be regarded as two overlapping Weyl points with opposite chirality, which can be separated in momentum space by applying a magnetic field. The separated Weyl nodes distribute along the field and the distance between them is proportional to the magnitude of magnetic field. The charge "pumping" effect between the Weyl nodes gives rise to the negative LMR in these TSMs, including both Dirac semimetals and Weyl semimetals.

In order to confirm the Dirac semimetal state of NdSb in the AFM state, the MR measurements are carried out to detect possible chiral-anomaly-induced negative LMR. The measurement geometry is illustrated in Fig. 4(a). $\theta=0$ means that the magnetic field (*H*) is applied perpendicular to the electric current (*I*), and $\theta=90°$ means that *H* is parallel to *I*. As shown in the inset of Fig. 4(a), the negative LMR is indeed observed in sample A at $\theta=90°$. Detailed MR data taken at different angles around $\theta=90°$ are presented in Fig. 4(b). The maximum negative LMR (~ -33%) appears at about 1.2 T for $\theta=90°$. When *H* is misaligned with respect to *I* (i.e., $\theta\neq90°$), the negative LMR gradually decays and finally vanishes for larger misalignment.

As shown in Fig 4(c), the negative LMR observed in NdSb is suppressed with increasing temperature and vanishes at 10 K, which is different from the behavior of magnetic scattering mechanism, i.e., the maximum MR appearing at the transition temperature. Thus the possibility of magnetic scattering mechanism can be ruled out.

In Dirac semimetals, the conductivity caused by chiral anomaly can be described using a semi-classical approximation formula [29]: $\sigma_\text{chiral} = C_\text{W}B^2$. Here $C_\text{W}$ is the chiral coefficient which is proportional to $v_\text{F}^3 \tau_a/(T^2 + \Delta E^2/\pi^2)$, where $v_\text{F}$ is the Fermi velocity near the Dirac points, $\tau_a$ is the inter valley scattering time, and $\Delta E$ is the measured chemical potential relative to the energy of Dirac points. In combination with the conventional conductivity ($\sigma_\text{N}$) and the conductivity from quantum interference effect ($\sigma_\text{Q}$), quantitative analyses can be carried out using the formula [16]: $\sigma(B) = (1 + C_\text{W}B^2) \cdot \sigma_\text{Q} + \sigma_\text{N}$, $\sigma_\text{Q} = \sigma_0 + a\sqrt{B}$ and $\sigma_\text{N}^{-1} = \rho_0 + A \cdot B^2$, where $\sigma_0$ is the zero-field conductivity, $a$ and $A$ are coefficients. For $\sigma_\text{Q}$, $a$>0 represents weak localization (WL) and *a*<0 represents weak anti-localization (WAL). The negative LMR of sample A is fitted based on the WL effect (here we set $C_W=0$) and chiral anomaly effect (in this case, we set *a*=0), respectively. As shown in Fig. 4(d), the fitting using WL model has a large mismatch with the experimental data, while the fitting using chiral anomaly model agrees well with the experimental data, indicating that the WL effect has a negligible contribution to the negative MR and the chiral anomaly could be the origin of negative LMR.

The chiral anomaly behavior can be quantitatively represented by chiral coefficient $C_W$. If the Weyl points are far away from the Fermi surface, the chiral anomaly induced negative MR will be too weak to be observed. The resultant $C_W$ for sample A shows angular and temperature dependence, as shown in Fig. 4(e). The maximum $C_W$ appears at $\theta=90°$, which is an important feature for the negative LMR induced by chiral anomaly. The rapid decrease in $C_W$ with temperature reveals the influence of thermal fluctuation, as shown in Fig. 4(f). The temperature-dependent $C_W$ below 6 K can be well fitted using $C_W \propto v_F^3 \tau_a/(T^2 + \Delta E^2/\pi^2)$ [29]. The resultant chemical potential is about 5 meV. It seems that the obtained chemical potential relative to the Dirac point is much smaller than the calculated value. We cannot give a good explanation on this divergence at present. One possible reason is the existence of some vacancy in actual samples. However, the relatively small chemical potential could contribute a lot to the observation of chiral-anomaly-induced negative LMR.

In large field regime, the MR becomes positive, which may be due to the electron-electron interaction from bulk bands. So besides the chemical potential from the energy of the Weyl points, the ingredient of bulk bands may also make the negative LMR very small, confined narrowly at low fields. Also note that the MR from normal bulk bands and chiral anomaly have different field dependence. For sample A, the chiral-anomaly-induced negative LMR is predominant below 1.3 T. While when the magnetic field is higher than 1.3 T, the bulk bands play an important role, leading to the gradual absence of negative LMR.

The inter valley scattering time at 2 K for sample A is estimated to be $\tau_a$=5.4×10$^{-10}$ s. The quasi-particle life time associated with the $\alpha'$ branch is estimated to be $\tau \simeq \hbar/E_F$=1.3×10$^{-13}$ s. We can see that the inter valley scattering time is much longer than the quasi-particle life time, which is a prerequisite to observe the negative LMR. The imbalance of population due to the inter valley scattering time can be estimated as $\Delta\mu = \hbar/\tau_a \simeq 1.2\times10^{-3}$ meV. These quantitative results are all characteristics of chiral anomaly in Dirac semimetals, which supports the existence of Dirac semimetal state in the low-field AFM state.

The observed negative MR can be well reproduced in other pieces of NdSb single crystal, as shown in Fig. 5(a) for sample B. In order to check whether or not the current jetting effect plays an important role in the negative LMR in NdSb, we perform the measurements of MR using multiple pairs of contacts on sample B, as shown in Fig. 5(b). The current electrodes $I_+$ and $I_-$ are sufficiently large to cover the shorter edges of crystal. At 2 K, the potential difference $V_i$ between the 4 pairs of nearest-neighbor contacts ($V_1$, $V_2$, $V_3$, $V_4$) are measured individually with sweeping the applied magnetic field. From Fig. 5(b) we notice that the four curves are nearly identical below 3.5 T, only displaying slight, nonsystematic deviations above 3.5 T. The nearly identical results provide solid evidence for the uniform current density throughout the crystal [31]. Thus it is reasonable to exclude the possibility of current-jetting in the observation of negative LMR in NdSb. The fitting results for sample B are shown in Figs. 5(c) and 5(d). The chemical potential relative to the Dirac point is about 5.5 meV for sample B, which is consistent with sample A.

The negative LMR for sample C exhibits some different features from the other two samples. The angle-dependent MR in Fig. 6(b) shows that the MR contains the contributions from weak anti-localization (WAL) effect, chiral anomaly and normal MR from bulk bands. Similar phenomenon has been widely observed in other TSMs. When the field is away from current direction, the negative LMR induced by chiral anomaly weakens and the normal MR becomes larger. The negative MR in sample C is similar to that reported in $WTe_2$. The observation of WAL depends on multiple factors, such as sample size, magnetic scattering, etc. The observed WAL for sample C might suggest an almost perfect condition for measuring large negative LMR, i.e., uniform current throughout the sample and completely parallel *H* and *I*. The temperature-dependent MR in Fig. 6(c) shows that the negative MR disappear at about 6 K. The resultant $C_W$ for sample C shows that the maximum $C_W$ appears at $\theta=90°$, as shown in Fig. 6(d). Using the same fitting, the chemical potential of sample C is estimated to be nearly zero. The small chemical potential turns out to be a crucial factor in observation of chiral anomaly induced negative LMR in Dirac and Weyl semimetals. Again, we notice that the actual chemical potential in the as-grown NdSb single crystal is significantly smaller than that determined from theoretical calculations, probably meaning that the as-grown NdSb single crystal has some vacancies. The situation is similar to that in $Bi_2Se_3$ topological insulator. However, it is mainly due to the advantage of the shifted chemical potential that we can detect the chiral anomaly induced negative LMR of NdSb.

For the FM state, the SdH oscillations are obtained after subtracting the non-oscillatory background from the MR data [Fig. 7(a)]. The oscillations above the AFM-FM transition (i.e., 14-38 T) are selected to perform the FFT analysis. Compared with the low-field data (Fig. 1), the FFT spectra under high-field show more discernible oscillation frequencies, i.e., 833 T ($\beta$), 1380 T ($\alpha'$) and 1633 T ($\gamma$) [Fig. 7(b)]. In Fig. 7(c), we give the enlarged view of the FFT spectra near the $\alpha$ branch. It is found that its frequency increases sharply with increasing temperature, reaching about 240 T at 16.1 K.

Upon the AFM-FM transition, it can be seen that the frequency of $\alpha$ branch is significantly shifted [Fig. 7(d)]. Another difference is the absence of low frequencies (the $\delta$ and $\varepsilon$ branches) in the high-field data. These differences suggest that the field-induced transition gives rise to a fundamental change in the topology of Fermi surface, i.e., reconstruction of Fermi surface. The Fermi surface reconstruction is consistent with the different band structures of AFM state and FM state. The vanishing electron pocket around K point could be related to the absence of low frequency branches. The shrunken bands at $\Gamma$ point is associated with the shift of $\alpha$ branch.

## IV. CONCLUSIONS

In summary, we find that NdSb is a Dirac semimetal in its AFM state. And the AFM spin orientation can be rearranged into an FM one, with the application of external magnetic field. A field-induced Fermi surface reconstruction is observed. Together with the recent discovery of magnetic Weyl semimetal state in $Sr_{1-y}Mn_{1-z}Sb_2$ (y, z < 0.1) [32], the discovery of AFM Dirac semimetal state in NdSb provides an important platform for exploring the interplay between Dirac fermions and magnetisms.

## ACKNOWLEDGMENTS

We thank B. J. Wieder for insightful discussions. This work was supported by National Key

R&D Program of China (Grant Nos. 2016YFA0300404 and 2017YFA0403600), the National Natural Science Foundation of China (Grant Nos. U1532267, 51603207, 11574288 and 11674327). Y.J. Wang and J. H. Yu contributed equally to this work.

*wkzhu@hmfl.ac.cn

†pili@ustc.edu.cn

‡zhangcj@hmfl.ac.cn

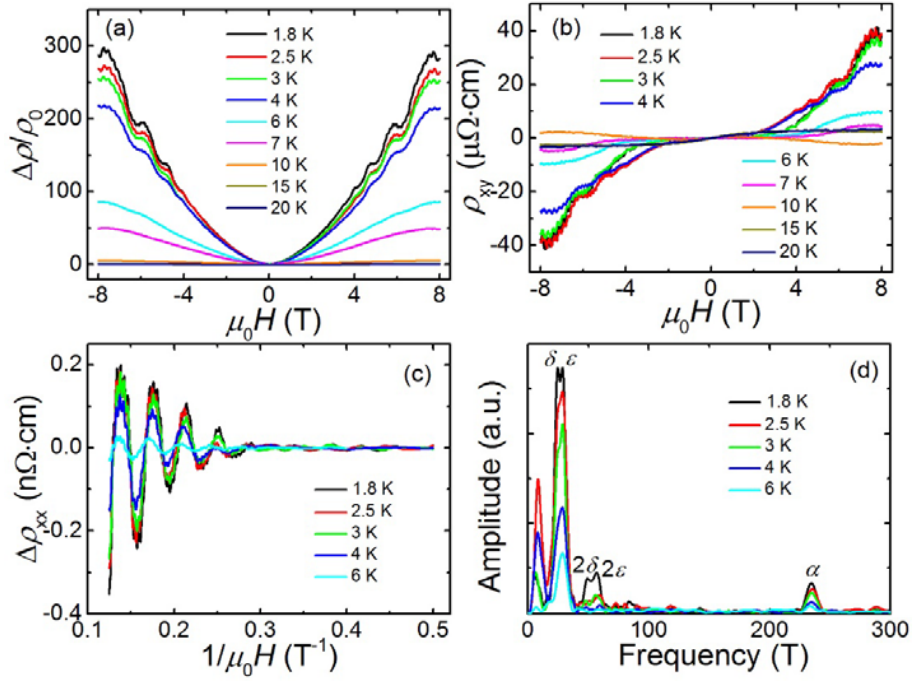

FIG. 1 (color online). (a) Magnetoresistance and (b) Hall resistivity $\rho_{xy}$ of NdSb in the range of 0-8 T, with the magnetic field applied perpendicular to the current and sample plane. (c) SdH oscillations vs $1/\mu_0 H$ in the range of 2-8 T. (d) FFT spectra of the SdH oscillations in (c).

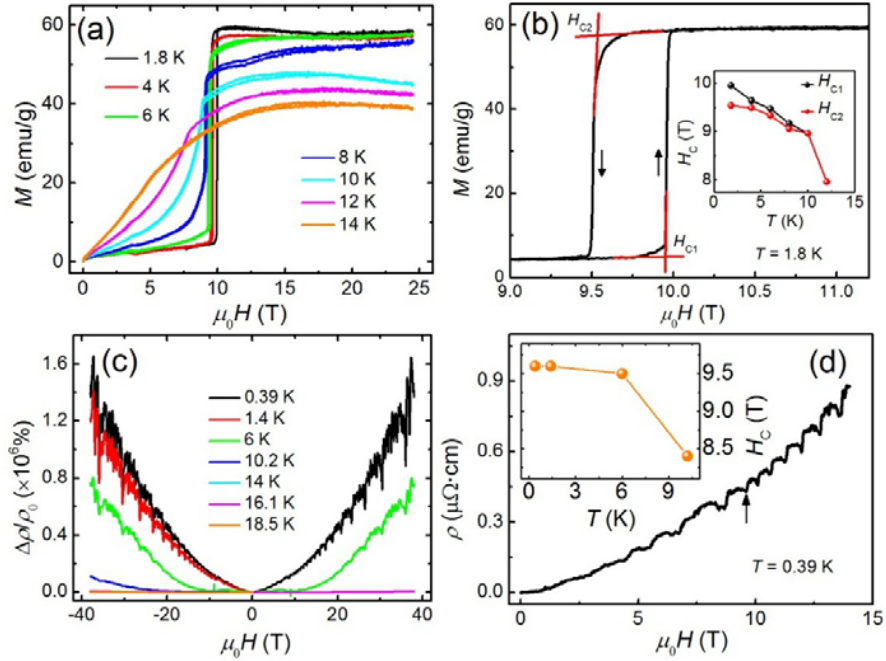

FIG. 2 (color online). (a) Magnetization vs magnetic field curves of the NdSb single crystal. (b) *M-H* curve between 9 T and 11 T at 1.8 K. Arrows indicate the direction of sweeping field. Inset: critical fields as a function of temperature. (c) High-field MR $\Delta\rho/\rho_0$ measured at different temperatures, with the magnetic field applied perpendicular to the current and sample plane. (d) The magnetoresistance in the range of 0-14 T at 0.39 K. Inset: $H_C$ as a function of temperature.

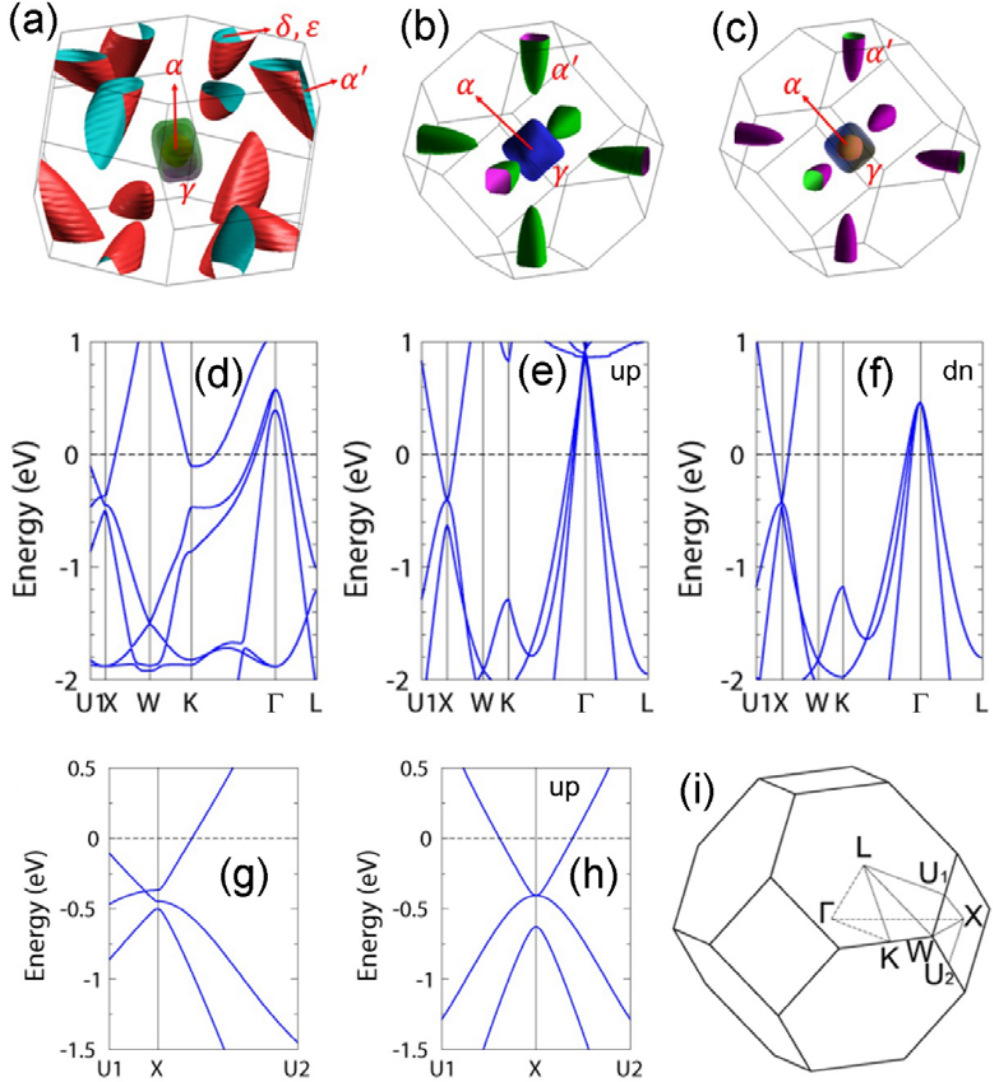

FIG. 3 (color online). The Fermi surfaces calculated for (a) AFM state (spin up and spin down are the same), and FM state with (b) spin up and (c) spin down. Band structures are plotted along different k-paths for (d)(g) AFM state and (e)(f)(h) FM state, respectively. Dashed lines indicate the Fermi level. The Fermi surface is characteristic of Fermi pockets centered at X, K and Γ. (i) High symmetry points in Brillouin zone. U1 and U2 are adjacent equivalent points for crystalline symmetry.

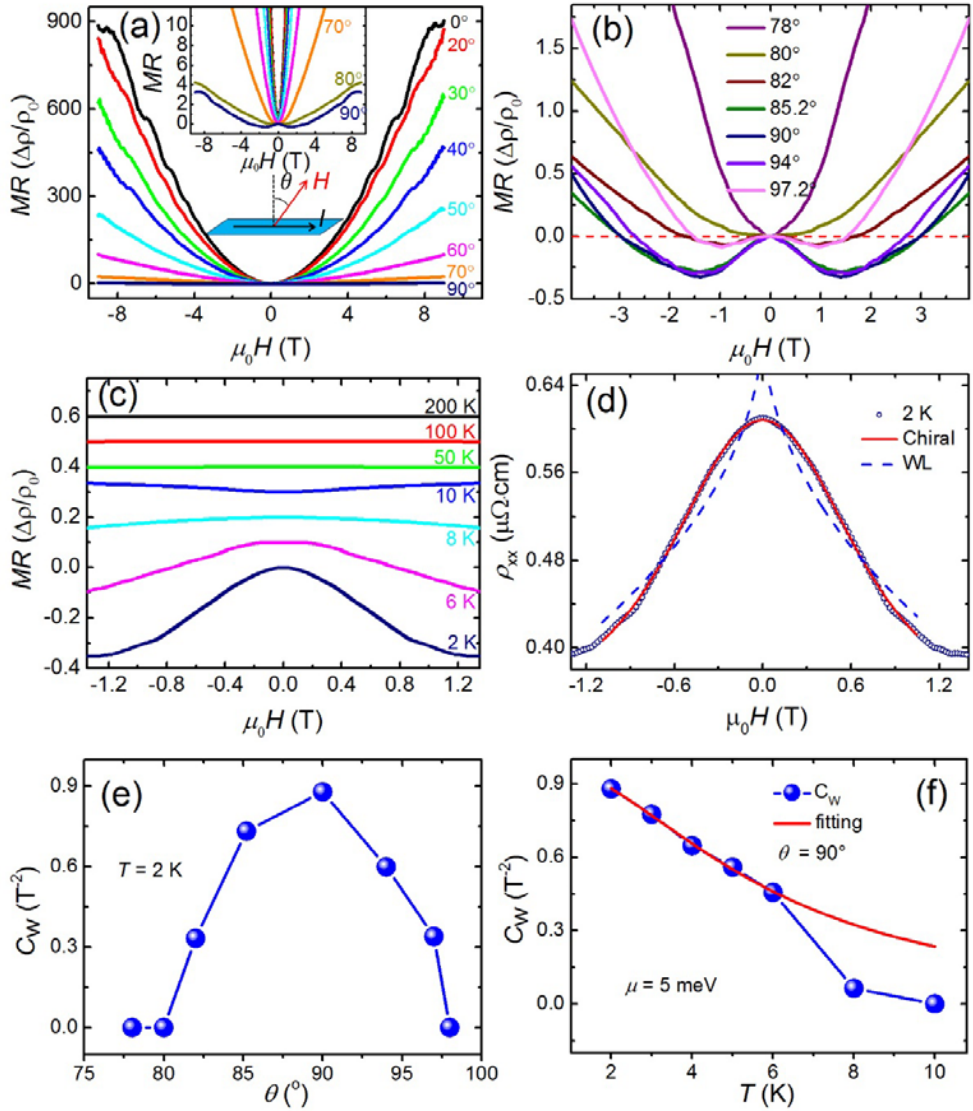

FIG. 4 (color online). (a) Low-field MR taken along different magnetic field directions at 2 K for sample A. The schematic diagram in (a) illustrates the measurement geometry. $\theta=90°$ means parallel magnetic field and electrical current. Inset: lower MR part shows negative MR at $\theta=90°$. (b) MR measured in different angles around $\theta=90°$ (from 78° to 97.2°). (c) MR measured at various temperatures with $\theta=90°$. The lines from 2 K to 200 K are shifted vertically by 0, 0.1, 0.2, 0.3, 0.4, 0.5 and 0.6, respectively. (d) Fittings for the negative MR data at 2K. Solid and dashed lines represent the fitted curves with chiral anomaly and WL formulas, respectively. (e) Angle-dependent and (f) temperature-dependent chiral coefficients. Solid curve represents the fitting results using the formula $C_W \propto v_F^3 \tau_v / (T^2 + \Delta E^2/\pi^2)$.

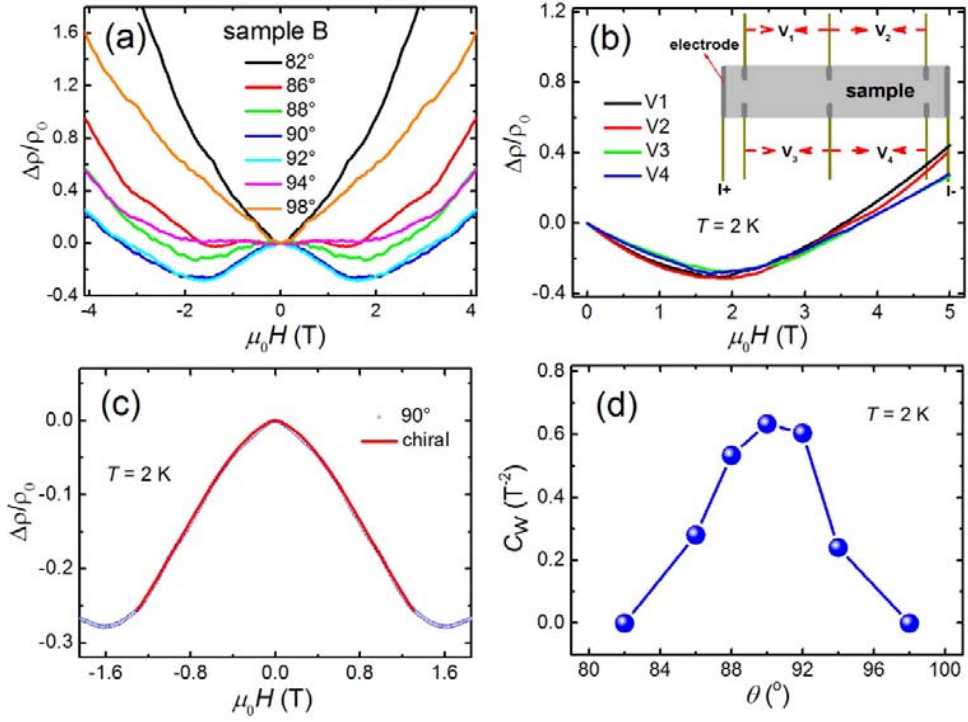

FIG. 5 (color online). (a) MR measured at different angles around $\theta=90°$ (from 82° to 98°) at 2 K for sample B. (b) MR taken between multiple pairs of contacts to rule out the current jetting effect. (c) Fitting results of the negative MR data at 2 K and $\theta=90°$. Solid line represents the fitted curve with chiral anomaly formula. (d) Angle-dependent chiral coefficients.

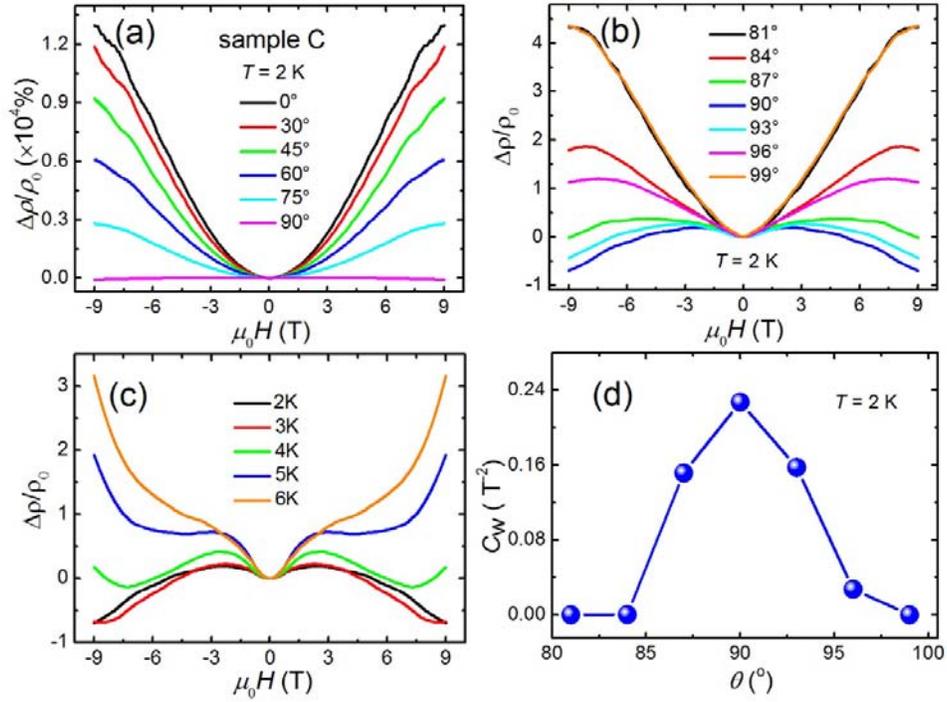

FIG. 6 (color online). (a) MR taken along different magnetic field directions at 2 K for sample C. (b) MR measured at different angles around $\theta=90°$ (from 81° to 99°). (c) MR measured at various temperatures with $\theta=90°$. (d) Angle-dependent chiral coefficients.

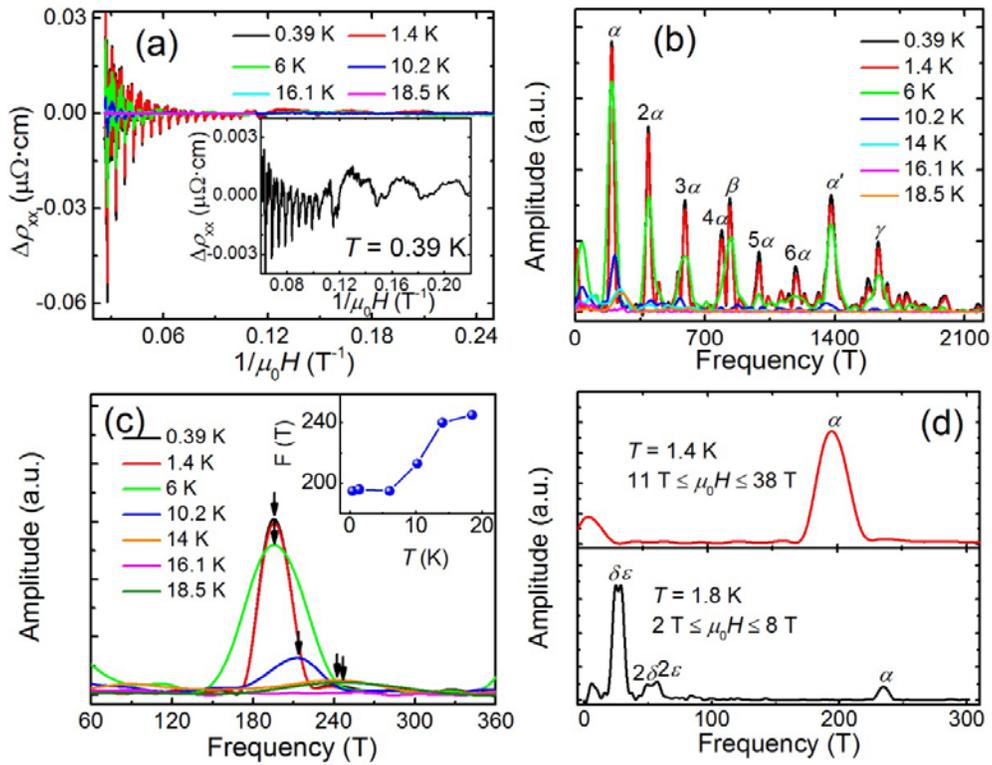

FIG. 7 (color online). (a) SdH oscillations vs $1/\mu_0 H$ of NdSb. Inset: oscillation at 0.39 K showing a transition at about 9.6 T. (b) FFT spectra of the SdH oscillations in (a). (c) FFT spectra of $\alpha$ branch at different temperatures. Inset: temperature dependence of the oscillation frequency of the $\alpha$ branch. (d) Comparison of FFT spectra for high field and low field in the frequency range of 0-300 T.